\documentclass[a4paper,11pt]{article}
\usepackage{cite}

\usepackage[margin=1.2in]{geometry}
\usepackage{graphicx}
\usepackage{amsmath, amssymb, amscd}
\usepackage{setspace} 
\newcommand{\vect}{\boldsymbol}

\begin{document}
\title{Moir\'e pattern as a magnifying glass for strain and dislocations in van der Waals heterostructures}
\author{Diana A. Cosma$^1$, John R. Wallbank$^1$, Vadim Cheianov$^1$,  Vladimir I. Fal'ko$^1$} 
\date{}

 \maketitle
\noindent{\small \textit{ 
$^1$Physics Department, Lancaster University, LA1 4YB, UK\\
} }

{\bf 
Two overlaying isostructural two-dimensional crystals with slightly different lattice constants, or two equivalent crystals deposited on each other with a slight misalignment, produce a long-range quasi-periodic structure \cite{Rayleigh,Righi} known as the moir\'e pattern. A rich library of moir\'e patterns has been found in graphene van der Waals (vdW) heterostructures on hexagonal catalysts \cite{diaye_prl_2006,marchini_prb_2007,sutter,sicot,batzill}, silicon carbide \cite{berger}, hexagonal boron nitride (hBN) \cite{xue,deker}, or deposited with a small misalignment angle on graphite \cite{rong}. 
For graphene electrons a moir\'e pattern translates into a superlattice with a period magnified from the atomic scale of the lattice constant $a$ by the large factor $M=[\theta^2 + \delta^2]^{-1/2}$ where $\delta\ll1$ is the lattice mismatch and $\theta\ll1$ is the misalignment angle.
We show that a similar factor magnifies the appearance of a dislocation in crystals composing a vdW heterostructure, in the form of a dislocation in the moir\'e superlattice, and also that a small uniaxial strain $w\sim1/M$ transforms a hexagonal moir\'e pattern into a quasi-one-dimensional incommensurate superlattice.}

\begin{figure}[tbhp]
\centering
\includegraphics[width=.6\textwidth]{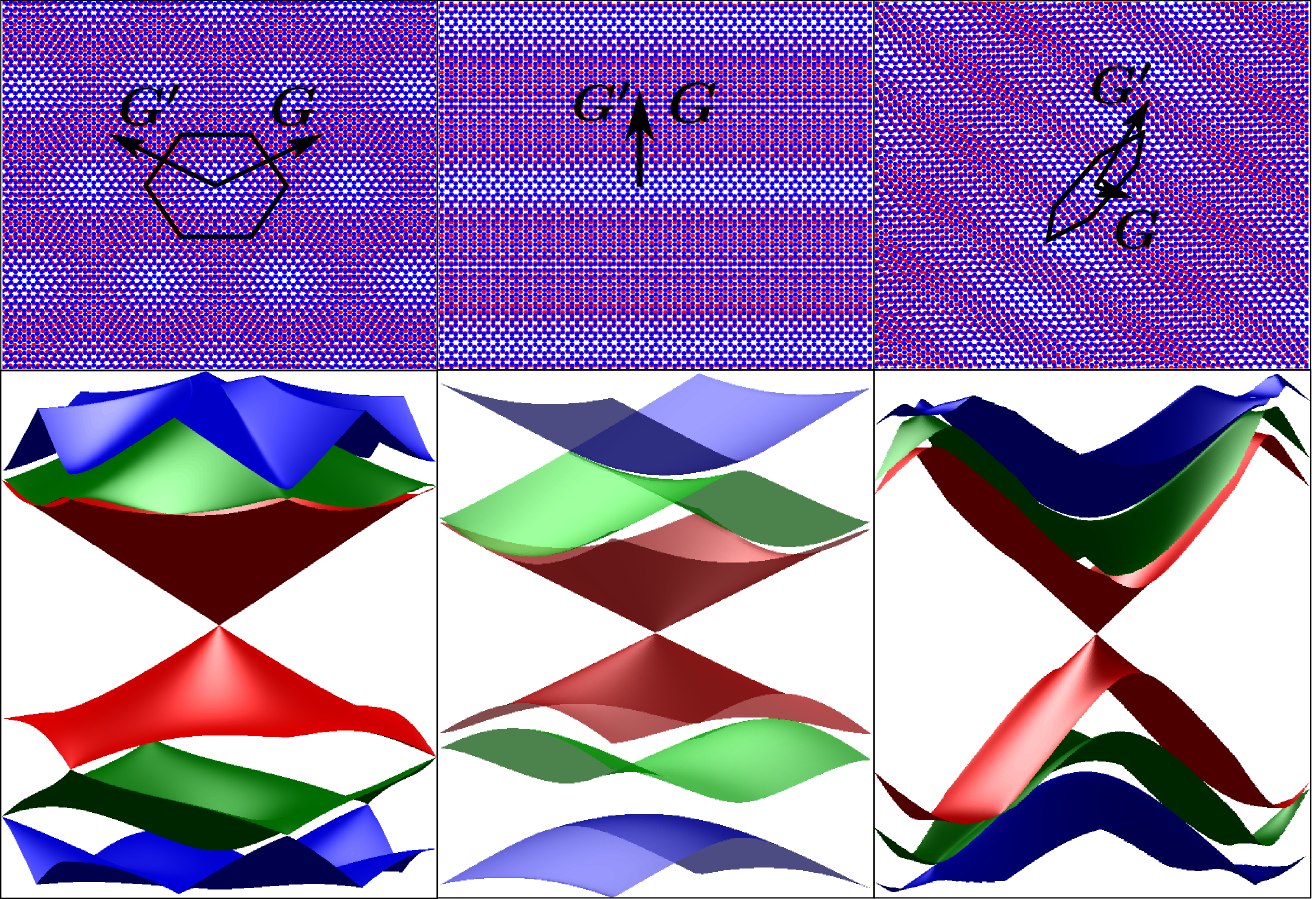}
\caption { 
Moir\'e superlattice in strained vdW heterostructures.
Top: Moir\'e patterns and corresponding moir\'e superlattice Brillouin zones. 
For graphene-hBN $\delta=1.8\%$, and we use $(w,\theta)=(0\%, 0^o)$, $(1.8\%, 0^o)$ and $(2\%, 0.92^o)$ (left to right).
Bottom: The corresponding band structures, calculated using Eq.~\eqref{eqn_moire_H} with $U_{i=0,1,3}\!=\!(v |\vect G'|/15) \{1/2,-1,- \sqrt 3/2 \}$.
 }
\label{fig1:moire}
\end {figure}

The upper panels of Fig.~\ref{fig1:moire} offers a geometrical illustration of how a small uniaxial strain, $w\ll1$ (either in graphene, or in the underlay) induces qualitative changes in the  moir\'e superlattice in a graphene-hBN heterostructure. The moir\'e superlattice of the unstrained heterostructure (l.h.s.~panel) transforms from hexagonal to oblique in the (r.h.s.~panel), taking a quasi-1D form at $w\sim\delta$ (central panel). The moir\'e lattice vectors ($m=0,1,\cdots5$),
\begin{align} \label{eq:moire_lattice}
\vect{A}_m= \hat{ M} \vect a_m,\qquad \hat M=\frac{1}{\delta'^2\!+\!\theta^2\!-\!w'^2}\begin{pmatrix} \delta'+(l_y^2-l_x^2)w' & \theta -2l_xl_y w'\\-\theta -2l_xl_y w'  &\delta'-(l_y^2-l_x^2)w'
 \end{pmatrix},
\end{align}
are magnified from graphene's lattice vectors $\vect a_m$ by the matrix $\hat{M}$, where, $\vect l=(l_x,l_y)$ is the principle axis of the strain tensor, $\delta'\!=\!\delta+w(1\!-\!\sigma)/2$, $w'\!=\!w(1\!+\!\sigma)/2$, and $\sigma$ is the Poisson ratio. 
Also, these expressions are quoted for strain applied to the substrate; for strain applied to the graphene layer we substitute $w\rightarrow-w$ and use $\sigma=0.165$ \cite{proctor}. The set of six vectors $\vect a_m$ is obtained by a $m\pi/3$ rotation of $\vect a_0=(a,0)$.

The perturbation produced by the moir\'e superlattice results in the formation of minibands for Dirac electrons in graphene \cite{park_natphys_2008,yankowitz,ortix, kinderman,ponomarenko,dean_nature_2013, wallbank,hunt_science_2013}.
The latter can be found by numerical diagonalisation of the moir\'e superlattice Hamiltonian,
\begin{align}
&\hat{H} =v\vect{p}\!\cdot\!\vect{\sigma}\!
+     
 \sum_{m} \left[ U_0 + (-1)^m \!\left(iU_3 \sigma_3 + U_1  \frac{\vect a_m}{a}\!\cdot\!\vect{\sigma}\right) \right] 
 e^{i \vect{G}_m\cdot  \vect{r}}e^{i \vect{G}_m\cdot \hat{M}\vect u_d(\vect{r})},\label{eqn_moire_H} 
\end{align}
where Pauli matrices $\sigma_{i=1,2,3}$ act on the electron amplitudes on graphene's $A$ and $B$ sublattices, and parameters $U_0$, $U_3$ and $U_1$ take into account a smooth potential, $A$-$B$ sublattice asymmetry and the modulation of $A$-$B$ hopping in graphene produced by the underlay \cite{wallbank}.
This moir\'e potential contains six harmonics characterised by Bragg vectors,
\begin{align}
\vect G_m = \frac{4\pi(\delta'^2+\theta^2-w'^2)}{\sqrt{3} a^2} \vect l_z\!\times\! \vect A_m, \nonumber
\end{align}
dependent on the misalignment angle and lattice mismatch, with an additional phase factor taking into account lattice deformations \cite{footnote_1}, $\vect u_d(\vect r)$, magnified by the matrix $\hat M$. Here, $\vect l_z$ is the unit vector normal to the 2D crystal plane.

 \begin{figure*}[tbhp]
 \centering
 \includegraphics[width=1\textwidth]{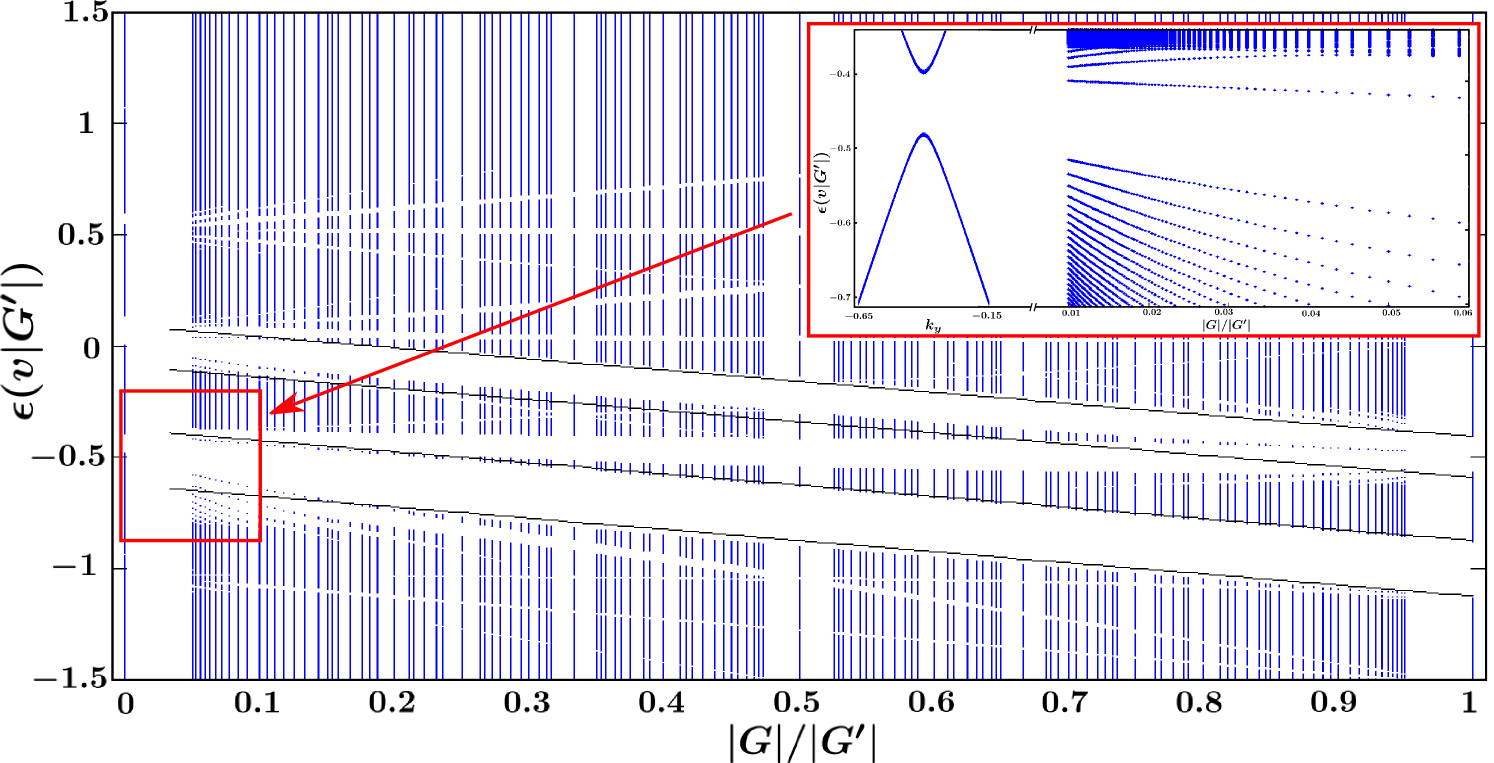}
\caption{ 
Strain-induced quasi-1D limit for moir\'e superlattices. 
Main panel: The spectral support of $\hat H$ calculated for a quasi-1D moir\'e superlattice with rational values of $|\vect{G}|/|\vect{G'}|$ for $k_x=0$ (same parameters $U_{0,1,3}$ as in Fig.~\ref{fig1:moire}). 
Black lines, marked at $\epsilon=(v\vect{G}\pm  |U_1+2U_3|)/2$ and $ \epsilon=(v\vect{G}+v\vect{G'}\pm|U_1+U_3|)/2$, show the approximate energies of band edges, calculated analytically for the two largest minigaps.
Inset: Miniband edges for the  $k_y$ dispersion for $k_x=|\vect{G}|/|\vect{G'}|=0$, compared with the spectral support for  $|\vect{G}|/|\vect{G'}|\leq0.06$ (left).}
\label{fig:butterfly}
 \end{figure*}
 
The anisotropy of the moir\'e superlattice in strained graphene-hBN heterostructures modifies the characteristic miniband spectrum for graphene electrons, as illustrated in the lower panels of Fig.~\ref{fig1:moire}. 
For the unstrained heterostructure (left) the spectrum exhibits a secondary Dirac point\cite{park_natphys_2008,yankowitz,ortix,ponomarenko,dean_nature_2013,wallbank} in the valence band. Strain, as low as $w\sim1-3\%$ (right), is sufficient for the secondary Dirac point to become obscured by overlapping spectral branches.
Most interestingly, a relatively weak strain,
\begin{align}
w_c&= 
\frac{(1-\sigma)\delta-\text{sign}(\delta)\sqrt{(1+\sigma)^2\delta^2+4\sigma \theta^2}}{2\sigma},\nonumber
\end{align}
transforms a hexagonal moir\'e superlattice into a quasi-1D lattice shown in the middle panel of Fig.~\ref{fig1:moire}. We find that such critical strain is independent of the orientation of the principal axes of the strain tensor and that, at $w=w_c$, all reciprocal lattice vectors are aligned along the same direction at an angle $\mathrm{arctan}(\frac{-\theta}{\delta+w_c})$ from $\vect l$.
In general such quasi-1D superlattice is characterised by two incommensurate lattice periods and a pair of two shortest Bragg vectors, $\vect G$ and $\vect G'$.
When the ratio $|\vect{G}|/|\vect{G'}|\!=p/q$ is rational, the electron spectrum is described in terms of plane waves with wavevector $\vect k=(k_x,k_y)$ where $| k_y|< |\vect G'|/q$ is counted along $\vect{G}$, and $|k_x|$, is unbound and counted perpendicular to $\vect{G}$.
The diagonalisation of Hamiltonian \eqref{eqn_moire_H} in the basis of Wanier wavefunctions corresponding to such wavenumbers returns a fractal spectrum  \cite{andre,simons_adv_app_math_1982} of ``bands'' and ``gaps'', for every value of $k_x$, as exemplified in Fig.~\ref{fig:butterfly}.
For a crystallographically aligned graphene-hBN heterostructure ($\theta=0$), strain $w=\delta=1.8\%$ exactly compensates the lattice mismatch. This results in the aligned reciprocal lattice vectors, with $|\vect{G}|/|\vect{G'}|\!=\!1$, as shown in Fig.~\ref{fig1:moire}~(centre). For a finite misalignment angle $\theta$ and $\vect l=\vect a_0/a$, the moir\'e pattern can be viewed as two superimposed incommensurate superlattices with an arbitrary ratio of periods, corresponding to
\begin{align}
\frac{|\vect{G}|}{|\vect{G'}|}&=\sqrt{\frac{f-1.14|\theta|\!}{f+1.14|\theta|\!}},\nonumber
\end{align}
where for the graphene-hBN pair 
$f\approx\!0.066 \sqrt{1 +1500\,\theta^2}-0.06$.
Hence, in general, the incommensurate structures possible in strained moir\'e superlattices have periods with $|\vect G|/ |\vect{ G' }|$ spanning across a considerable range of, $0<|\vect G|/ |\vect{ G' }|<1$.

Fractal spectra of electrons, calculated for $k_x\!=\!0$, and various commensurate ratios $|\vect{G}|/|\vect{G'}|\!=p/q$ (controlled by the misalignment angle) are shown in Fig.~\ref{fig:butterfly}. The hierarchy of ``gaps'' and bands in this spectrum, illustrated in the inset for $p/q=0$, can be understood by using the $\vect k\cdot \vect p$ theory for the dispersion of electrons at a simpler fraction (such as $p/q=0$; $1$; or $1/2$) and including incommensurability in the form of an additional perturbation.
For example, the vicinity of $p/q=0$ can be described by using the parabolic dispersion of electrons near the band edges of a spectrum evaluated for finite $\vect{G}'$ (see inset in Fig.~\ref{fig:butterfly}), where as an additional slow-varying perturbation produces characteristic size-quantised levels near minima/maxima of the potential oscillating with wavevector $\vect{G}$.
The same approach enables us to approximate the edges of the largest gaps in the spectrum, shown by black lines. Note that for all $k_x\neq0$ such gaps appear at different energies, hence the overall fractal spectrum is not gapped.

Up to now, we have considered the effect of homogeneous strain on moir\'e superlattices. Beyond that, any defect-induced smooth displacement field, $\vect u_d(\vect r)$, also affects the moir\'e superlattice geometry, and a smoothly varying local coordination between graphene and the substrate atoms would reflect such deformations in the Hamiltonian \eqref{eqn_moire_H} of graphene electrons.
To follow how the actual lattice deformations can be related to the deformations of the moir\'e superlattice, we compare the points of constant phase (which are analogous to lattice ``sites'') in the Fourier harmonics included in Eq.~\eqref{eqn_moire_H} with ($\vect R_d$) and without ($\vect R$) deformations.
Then, the displacement field in the moir\'e superlattice, $\vect U_d\left(\vect R_d\right)\!=\!\vect R_d-\vect R$, describes the position of a particular moir\'e lattice ``site'' in the presence of a defect, relative to its position in the absence of a defect. Using Eq.~\eqref{eqn_moire_H}, we find that
\begin{equation}
\vect U_d\left(\vect r\right) =- \hat{ M}\vect u_d(\vect r),
 \label{eqn_defect}
\end{equation}
which enables us to match extended and topological defects in the top/bottom layer of the heterostructure to a partner defect in the moir\'e superlattice, which appears at a scale magnified by $\hat M$.

 \begin{figure*}[tbhp]
\centering
\includegraphics[width=1\textwidth]{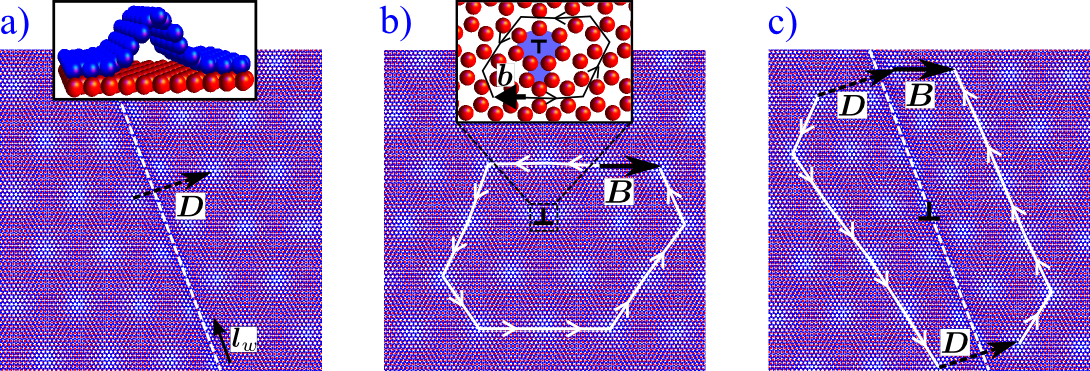}
\caption{
Moir\'e pattern in vdW heterostructure with extended defects in one of the crystal layers.
(a) VdW heterostructure with a wrinkle present in the graphene layer. 
(b) A dislocation in the underlay generates a dislocation in the moir\'e pattern.
(c) Dislocation in the underlay and a wrinkled graphene layer. }
\label{fig1:moire_magnify}
\end{figure*}

Below, we discuss in detail the two defect types shown in Fig.~\ref{fig1:moire_magnify}. Figure~\ref{fig1:moire_magnify}(a) illustrates a wrinkle in the top graphene layer centred at the dashed white line (oriented along unit vector $\vect l_w$).
The effect of such a wrinkle on the moir\'e superlattice is characterized by a vector $\vect D$ connecting moir\'e lattice ``sites'' on opposite sides of the wrinkle, measured at a distance larger than the wrinkle's width. 
Then, the shift 
\begin{align}
\vect d\equiv\!\int_{\vect D}\!  d\!s \frac{d\vect u_d\!(\vect r(s))}{d\!s},\nonumber
\end{align}
in atomic positions of carbon atoms with respect to their position in flat graphene, can be found using
\begin{gather}
\vect d\!=-\hat {M}^{-1}(\vect D+n_0\vect A_0 +n_1\vect A_1).
\label{eq:wrinkles}
\end{gather}
Here, integers $n_{0}$ and $n_{1}$ account for the fact that $\vect D$ is only measured up to a vector from the moir\'e superlattice and should be used to choose the smallest possible $\vect d$. 
Equation \eqref{eq:wrinkles} is also valid if a discontinuity in the moir\'e superlattice is due to a step edge in the underlay~\cite{coraux_nano_lett_2008}. In this case, $\vect u_d(\vect r)$ would include both the effect of curving the graphene flake over the step edge, as well as the shift in stacking characterised by the different atomic layers in the underlay.

In Figure~\ref{fig1:moire_magnify}(b) we illustrate how a dislocation in the substrate, characterised by Burgers vector $\vect b$, is reflected by a dislocation in the moir\'e superlattice, with burgers vector $\vect B$:
\begin{gather}
 \vect{b} \equiv \oint d\!s \frac{d\vect{u}_d(\vect{r}(s))}{ds}=-\hat {M}^{-1} \vect B, \qquad \vect B\equiv \oint d\!s \frac{d\vect{U}_d(\vect{r}(s))}{ds}.
\label{eq:burgers}
\end{gather}
A dislocation in graphene with Burger's vector $-\vect b$ will have the same effect on the moir\'e superlattice.
Interestingly, the moir\'e pattern analysis enables one to determine the presence of a dislocation even in a wrinkled heterostructure, as illustrated in Fig.~\ref{fig1:moire_magnify}(c).

\begin{figure}[tbhp]
\centering
\includegraphics[width=.7\textwidth]{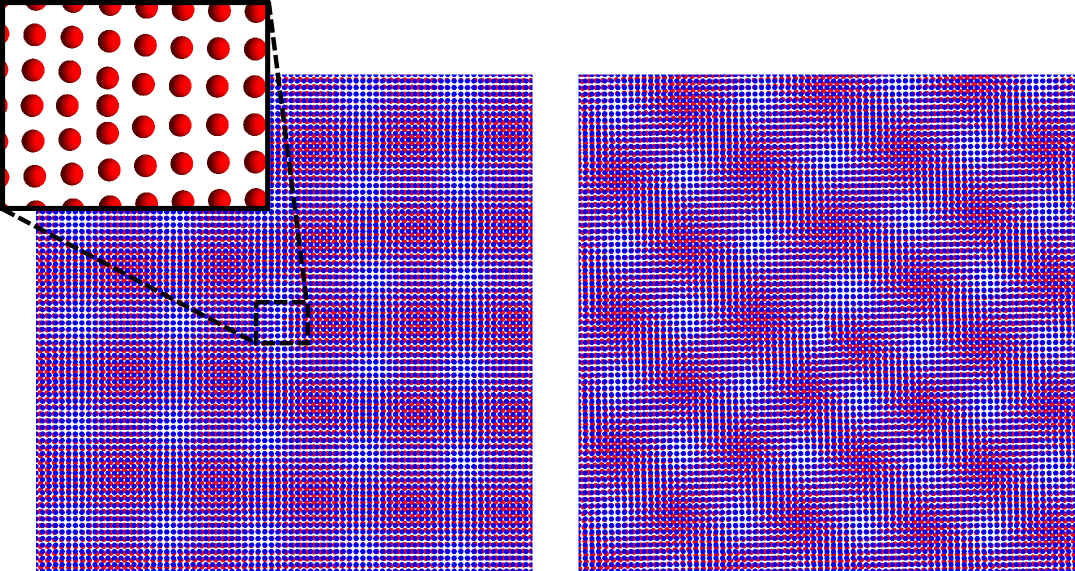}
\caption {
Moir\'e patterns for vdW heterostructure of square lattice crystals with dislocations. Left: $\delta=0.072$ and $\theta=0$. Right: $\delta=0.05$ and $\theta=0.05$.}
\label{fig1:sq_lattice}
\end {figure}

In conclusion, we have shown that moir\'e superlattices characteristic for graphene heterostructures with hexagonal crystals, or graphene deposited on a substrate with an hexagonal surface layer, magnify the length-scale of deformations in either of the involved crystals, by generating deformations and defects in the moir\'e pattern. This result, discussed in detail for heterostructures of hexagonal crystals, remains qualitatively the same for other crystalline lattices. For example, in Fig.~\ref{fig1:sq_lattice}, we show that a dislocation in one of two square lattice crystals in a heterostructure generates a clearly identifiable dislocation in the square moir\'e superlattice.

\subsection*{Methods}
Due to a larger separation $d\gg a$ between graphene and its substrate as compared to  the lattice constant of graphene, and also due to their weak coupling, 
Dirac electrons in graphene's bands experience the effect of the substrate only via Bragg scattering processes with momentum transfer less than $d^{-1}$ \cite{ortix,kinderman,wallbank}. This determines the set of Fourier harmonics included in Hamiltonian \eqref{eqn_moire_H}, with only those differences between reciprocal lattice vectors in graphene, $\vect G^G_m$, and hBN, $\vect G^{hBN}_m$, that satisfy the condition $|\vect G^{hBN}_m|d\ll1$.
The Hamiltonian \eqref{eqn_moire_H} includes only terms which are symmetric under the in-plane spatial inversion, which assumes that one of the sublattices (boron or nitrogen) of the hBN layer affects the graphene Dirac electrons much stronger than the other, making the substrate effectively inversion symmetric.
Taking into account weak inversion asymmetry opens gaps in graphene's miniband spectrum, however this does not affect the qualitative changes in the moir\'e pattern and the resulting minibands induced by deformations.

To incorporate the effect of deformations in the Hamiltonian \eqref{eqn_moire_H} we assumed that its form, locally, is determined by a local coordination of atoms in the two layers, displaced by $\vect u_d(\vect r)$ from their positions in a non-deformed heterostructure. Such a shift changed the local phase of harmonics corresponding to the oscillations of the three local symmetry-breaking perturbations created by hBN atoms for electrons in graphene. We use this additional phase shift to identify formal ``sites'' of the moir\'e superlattice, as points where $\vect G_m \cdot \left[\vect r+\hat M \vect u_d(\vect r)\right]=2\pi N$. To describe the effect of extended/topological defects in the 2D crystal lattice, we incorporate the long-range (distances $\gg a$) deformations caused by them in $\vect u_d(\vect r)$, used in Eq.~\eqref{eqn_moire_H}.

To calculate the spectra of $\hat{H}$ shown in Figs.~\ref{fig1:moire}-\ref{fig:butterfly}, we used zone folding for Dirac electrons into a mini Brillouin zone determined by the moir\'e superlattice, which is simplified by the lack of intervalley scattering (the latter would require a too large momentum transfer). Then we numerically diagonalize the resulting Heisenberg matrix for each momentum point in the mini Brillouin zone.
For numerical diagonalisation, we choose phenomenological parameters  $U_0\!=\!v|\vect G'|/30$, $U_1\!=\!-v|\vect G'|/15$, and $U_3\!=\!-v|\vect G'|\sqrt{3}/30$, whose relative size corresponds to two specific microscopic models \cite{wallbank}.

\section*{Acknowledgements}
This work was supported by EC FP7 Graphene Flagship project CNECT-ICT-604391, ERC Grant Hetero2D, and the Royal Society Wolfson Research Merit Award. We are grateful to H. Schomerus and A.~Geim for useful discussions.

\end{document}